# Lie symmetry and the Bethe ansatz solution of a new quasi-exactly solvable double-well potential

M. Baradaran and H. Panahi[1]

*Department of Physics, University of Guilan, Rasht 41635-1914, Iran*

**Abstract**

In this paper, we study the Schrödinger equation with a new quasi-exactly solvable double-well potential. Exact expressions for the energies, the corresponding wave functions and the allowed values of the potential parameters are obtained using two different methods, the Bethe ansatz method and the Lie algebraic approach. Some numerical results are reported and it is shown that the results are in good agreement with each other and with those obtained previously via a different method.

## 1. Introduction

A quantum mechanical system is exactly solvable (ES) if all the eigenvalues and corresponding eigenfunctions can be determined exactly through algebraic means. These quantum systems play an important role in various branches of physics. However, such systems are rare and the Schrödinger equation cannot be solved exactly to obtain the whole spectrum except for a limited number of potentials, such as the harmonic oscillator, Coulomb and Pöschl-Teller potentials [1-5]. A review of early works in this area can be found in Refs. [6-8]. In contrast, a quantum system is called quasi-exactly solvable (QES) if only a finite part of the spectrum can be found exactly [9-12]. During the last decades, the QES models have received a great deal of attention because of their wide applications in quantum mechanics [13-21]. These models are distinguished by the fact that their infinite-dimensional Hamiltonian can be reduced to a block diagonal matrix with at least a finite-dimensional block, in which its eigenvalues and eigenfunctions can always be determined by diagonalizing the corresponding matrix. On the other hand, during the last decades, a great deal of attention has been given to the study of the Schrödinger equation with QES double-well potential (DWP) including the quartic potential [22], the sextic potential [23] and the Razavy potential [24]. In the literature, there are two distinct approaches for investigating the QES systems: the Lie algebraic approach [9-11] and the analytical approach [12,15] which is based on the Bethe ansatz method (BAM). The interested reader is referred to Refs. [4, 25-29] and references therein for more detailed information regarding the application of the wave function ansatz method in physical problems. In this paper, applying the analytical approach of quasi-exact solvability, we investigate the Schrödinger equation for a new type of one-dimensional QES DWP proposed by Chen et.al. [17]. They studied the problem and obtained solutions of the first two states by using two methods, the confluent Heun functions and the Wronskian method [17]. Within the present study, through the BAM, we are going to extend the results of Ref. [17] by finding general exact expressions for the energies, the wave functions and the special constraints on the potential parameters. Also, we solve the same problem using the Lie algebraic approach and illustrate how the relation with the $sl(2)$ Lie algebra underlies the quasi-exact solvability of the DWP model, as well as we show that the results of the two methods are consistent.

[1] Corresponding author E-mail: t-panahi@guilan.ac.ir

This paper is organized as follows: In section 2, we introduce the QES DWP and solve the corresponding Schrödinger equation using the BAM. Also, the closed form expressions for the energies and the wave functions are obtained in terms of the roots of the Bethe ansatz equations. In section 3, we solve the same problems using the Lie algebraic approach and demonstrate that the system possesses a hidden $sl(2)$ algebraic symmetry which is responsible for quasi-exact solvability. Some numerical results, obtained by the BAM and QES methods are reported and discussed in section 4. Finally, in section 5, we present the conclusions.

## 2. The analytical method based on the Bethe ansatz for the QES DWP

We begin with the one-dimensional three parameter QES DWP proposed by Chen et al. [17] as

$$V(x) = \frac{v_1}{\cosh^2(x)} + \frac{v_2}{1 + g\cosh^2(x)} + \frac{v_3}{(1 + g\cosh^2(x))^2}, \tag{1}$$

which under the condition $g \gg 1$, becomes the Manning potential [30]

$$V_{Manning}(x) = v_4 \sec h^2(x) + v_5 \sec h^4(x), \tag{2}$$

$$v_4 = v_1 + \frac{v_2}{g}, \quad v_5 = \frac{v_3}{g^2}.$$

In 1935, Manning used this symmetric double minima potential to study the vibrational normal modes of the ND3 and NH3 molecules [30]. This application is possible because the nitrogen atom in these molecules has two equilibrium positions on either side of the D2 and H2 planes. In the following, we show that under certain constraints on the potential parameters $v_1$, $v_2$ and $v_3$, a finite number of the energy eigenvalues and eigenfunctions of the corresponding Schrödinger operator can be obtained exactly in explicit form. In Fig. 2, we draw the potential (1) for the allowed values of potential parameters. In atomic units ($m = \hbar = c = 1$), the Schrödinger equation with potential (1) is written as

$$\left(-\frac{d^2}{dx^2} + \frac{v_1}{\cosh^2(x)} + \frac{v_2}{1 + g\cosh^2(x)} + \frac{v_3}{(1 + g\cosh^2(x))^2}\right)\psi(x) = E\psi(x). \tag{3}$$

Chen et al. [17] studied this problem earlier and presented the results of the first two states in terms the confluent Heun functions. Here, we intend to extend their results by calculating the general exact solutions of the problem in terms of the roots of a set of algebraic equations within the framework of the Bethe ansatz. For this aim, making the change of variable $z = -\sinh^2(x)$ and taking the following ansatz for the wave function

$$\psi(x) = (\cosh(x))^{\frac{1}{2}\left(1 + \sqrt{1 - 4v_1}\right)} (1 + g\cosh^2(x))^{\frac{1}{2}\left(1 - \sqrt{1 + \frac{v_3}{1+g}}\right)} \phi(z), \tag{4}$$

we obtain

$$H\phi(z) = 0,$$

$$H = \left(z^3 - (\lambda + 1)z^2 + \lambda z\right)\frac{d^2}{dz^2} + \left((\gamma + \delta + \varepsilon)z^2 - (\lambda\gamma + \gamma + \lambda\delta + \varepsilon)z + \lambda\gamma\right)\frac{d}{dz} + (\alpha\beta z - \sigma), \tag{5}$$

where

$$\alpha = \frac{1}{4}\left(3 + \sqrt{1 - 4v_1} - 2\sqrt{1 + \frac{v_3}{1+g}} + 2\sqrt{-E}\right),$$

$$\beta = \frac{1}{4}\left(3 + \sqrt{1 - 4v_1} - 2\sqrt{1 + \frac{v_3}{1+g}} - 2\sqrt{-E}\right),$$

$$\gamma = \frac{1}{2}, \qquad \delta = 1 + \frac{1}{2}\sqrt{1 - 4v_1}, \qquad \lambda = \frac{1+g}{g}, \tag{6}$$

$$\sigma = \frac{\lambda}{4}\left(\frac{1}{2} + \frac{1}{2}\sqrt{1 - 4v_1} + \frac{1}{\lambda} - \frac{1}{\lambda}\sqrt{1 + \frac{v_3}{1+g}} - v_1 - \frac{v_2}{g\lambda} - \frac{v_3}{g^2\lambda^2} + E\right),$$

$$\varepsilon = \alpha + \beta + 1 - \gamma - \delta.$$

In order to apply the BAM to the present problem, we suppose that (5) has degree $n$ polynomial solution as (the Bethe ansatz)

$$\phi_n(z) = \begin{cases} \prod_{k=1}^{n}(z-z_k) & n \neq 0 \\ 1 & n = 0 \end{cases}, \tag{7}$$

with undetermined, distinct roots $z_k$ which are none other than the wave function nodes. Substituting Eq. (7) into Eq. (5), and after some algebra, we obtain

$$\frac{\lambda}{4}\left(\frac{1}{2}+\frac{1}{2}\sqrt{1-4v_1}+\frac{1}{\lambda}-\frac{1}{\lambda}\sqrt{1+\frac{v_3}{1+g}}-v_1-\frac{v_2}{g\lambda}-\frac{v_3}{g^2\lambda^2}+E\right) =$$
$$(n(n-1)+n(\gamma+\delta+\varepsilon)+\alpha\beta)z$$
$$+\sum_{k=1}^{n}\frac{1}{z-z_k}\left(\left(z_k^3-(\lambda+1)z_k^2+\lambda z_k\right)\sum_{j\neq k}\frac{2}{z_k-z_j}+\left((\gamma+\delta+\varepsilon)z_k^2-(\lambda\gamma+\gamma+\lambda\delta+\varepsilon)z_k+\lambda\gamma\right)\right) \tag{8}$$
$$+(2(n-1)+\gamma+\delta+\varepsilon)\sum_{k=1}^{n}z_k-n(n-1)(\lambda+1)-n(\lambda\gamma+\gamma+\lambda\delta+\varepsilon).$$

Comparing the left and right hand sides of Eq. (8), we obtain the following relations

$$-\sum_{j\neq k}^{n}\frac{2}{z_k-z_j}-\frac{\gamma}{z_k}-\frac{\delta}{z_k-1}-\frac{\varepsilon}{z_k-\lambda}=0, \tag{9}$$

$$\frac{1}{16}\left(3+\sqrt{1-4v_1}-2\sqrt{1+\frac{v_3}{1+g}}\right)^2+\frac{E}{4}=-n(n-1)-n(\gamma+\delta+\varepsilon), \tag{10}$$

$$-\left(\frac{\lambda}{4}\left(\frac{1}{2}+\frac{1}{2}\sqrt{1-4v_1}+\frac{1}{\lambda}-\frac{1}{\lambda}\sqrt{1+\frac{v_3}{1+g}}-v_1-\frac{v_2}{g\lambda}-\frac{v_3}{g^2\lambda^2}+E\right)\right)$$
$$=-(2(n-1)+\gamma+\delta+\varepsilon)\sum_{k=1}^{n}z_k+n(n-1)(\lambda+1)+n(\lambda\gamma+\gamma+\lambda\delta+\varepsilon), \tag{11}$$

for the roots $z_k$ (the so-called Bethe ansatz equation), the energy eigenvalues and the restrictions on the parameters of the potential, respectively. Now, Let us look at the Bethe ansatz equations (9) from a different point of view. It is not difficult to understand that the set of algebraic equations (9) can be interpreted as the equilibrium condition of an analogue mechanical system consisting of $n$ particles with coordinates $z_k$ moving in the interaction potential

$$U(z_1,...,z_n)=-2\sum_{j\neq k}^{n}\ln|z_k-z_j|-\sum_{k=1}^{n}(\gamma\ln z_k+\delta\ln(z_k-1)+\varepsilon\ln(z_k-\lambda)). \tag{12}$$

In other words, the $n$ equations $\partial U/\partial z_k$ ($k=1,2,...,n$) describing the equilibrium distribution of the particles, coincide with Eq. (9). The first term in this expression is none other than the Coulomb repulsion of the particles while the three terms in the parenthesis correspond to the interactions of each individual particle with the force centers located at the origin, 1 and $\lambda$, respectively. Therefore, from the point of view of classical mechanics, the problem of finding the roots of the set of algebraic Bethe ansatz equations (9) is equaivalent to the problem of finding equilibrium positions of a system of classical Coulomb particles in the interaction potential $U(z_1,...,z_n)$. In the following, to clarify how the method works, we study the ground, the first and the second excited states of the model in more detail. For $n=0$, from Eqs. (10) and (4), we have the following relations

$$\frac{1}{16}\left(3+\sqrt{1-4v_1}-2\sqrt{1+\frac{v_3}{1+g}}\right)^2+\frac{E}{4}=0, \tag{13}$$

$$\psi_0(x)=(\cosh(x))^{\frac{1}{2}\left(1+\sqrt{1-4v_1}\right)}(1+g\cosh^2(x))^{\frac{1}{2}\left(1-\sqrt{1+\frac{v_3}{1+g}}\right)},$$

for the ground state energy and the corresponding wave function, where the potential parameters from Eq. (11) satisfy

$$\frac{\lambda}{4}\left(\frac{1}{2}+\frac{1}{2}\sqrt{1-4v_1}+\frac{1}{\lambda}-\frac{1}{\lambda}\sqrt{1+\frac{v_3}{1+g}}-v_1-\frac{v_2}{g\lambda}-\frac{v_3}{g^2\lambda^2}+E\right)=0. \tag{14}$$

For $n=1$, by Eqs. (10) and (4), we obtain the first excited state energy and wave function as

$$\frac{1}{16}\left(3+\sqrt{1-4v_1}-2\sqrt{1+\frac{v_3}{1+g}}\right)^2+\frac{E}{4}=-(\gamma+\delta+\varepsilon),$$

$$\psi_1(x)=(\cosh(x))^{\frac{1}{2}\left(1+\sqrt{1-4v_1}\right)}(1+g\cosh^2(x))^{\frac{1}{2}\left(1-\sqrt{1+\frac{v_3}{1+g}}\right)}\left(-\sinh^2(x)-z_1\right), \tag{15}$$

respectively. In this case, the constraint on the potential parameters is given by

$$-\frac{\lambda}{4}\left(\frac{1}{2}+\frac{1}{2}\sqrt{1-4v_1}+\frac{1}{\lambda}-\frac{1}{\lambda}\sqrt{1+\frac{v_3}{1+g}}-v_1-\frac{v_2}{g\lambda}-\frac{v_3}{g^2\lambda^2}+E\right)$$
$$+(\gamma+\delta+\varepsilon)z_1+(\lambda\gamma+\gamma+\lambda\delta+\varepsilon)=0, \tag{16}$$

where the root $z_1$ of the wave function is obtainable from the Bethe ansatz equation (9) as

$$z_1=\frac{\lambda\delta+\lambda\gamma+\gamma+\epsilon\pm\sqrt{(\lambda-1)^2\gamma^2+2(\lambda-1)(\lambda\delta-\epsilon)\gamma+(\lambda\delta+\epsilon)^2}}{2\gamma+2\delta+2\epsilon}. \tag{17}$$

Analogusly, for the second excited state $n=2$, we have

$$\frac{1}{16}\left(3+\sqrt{1-4v_1}-2\sqrt{1+\frac{v_3}{1+g}}\right)^2+\frac{E}{4}=-2-2(\gamma+\delta+\varepsilon),$$

$$\psi_2(x)=(\cosh(x))^{\frac{1}{2}\left(1+\sqrt{1-4v_1}\right)}(1+g\cosh^2(x))^{\frac{1}{2}\left(1-\sqrt{1+\frac{v_3}{1+g}}\right)} \tag{18}$$
$$\times\left(\sinh^4(x)+(z_1+z_2)\sinh^2(x)+z_1z_2\right),$$

for the energy and wave function, respectively, where the potential parameters satisfy the constraint condition

$$-\frac{\lambda}{4}\left(\frac{1}{2}+\frac{1}{2}\sqrt{1-4v_1}+\frac{1}{\lambda}-\frac{1}{\lambda}\sqrt{1+\frac{v_3}{1+g}}-v_1-\frac{v_2}{g\lambda}-\frac{v_3}{g^2\lambda^2}+E\right)$$
$$+(2+\gamma+\delta+\varepsilon)(z_1+z_2)+2(\lambda+1)+2(\lambda\gamma+\gamma+\lambda\delta+\varepsilon)=0, \qquad (19)$$

and the roots $z_1$ and $z_2$ are obtainable from the Bethe ansatz equations (9) as

$$(z_1,z_2)=\begin{cases}(9.714045208,\ 0.2859547921)\\ (0.6953879418,\ 0.1092848103)\\ (7.229242571-4.010375187i,\ 7.229242571+4.010375187i).\end{cases} \qquad (20)$$

Here, we have taken the parameters $v_1=0.09$, $v_3=10$ and $g=1/4$. In Table 1, we have reported and compared our numerical results for the first four states. As an additional comment on the treatment of the wave functions, from Eqs. (13), (15) and (18) it can be seen that except the ground state wave function, the higher excited states are mathematically meaningless for large $x$, since as $x\to\pm\infty$, we have $|z|=\sinh^2(x)\to\infty$. This property can be explained well by the asymptotic behavior of the Heun function which is only convergent within the circle $|z|<1$. The interested reader is referred to Ref. [17] for details of the problem.

## 3. The Lie algebraic approach for the QES DWP

In the previous section, we have investigated the Schrödinger equation for the QES DWP and obtained the general exact solutions of the system within the framework of the Bethe ansatz. In this section, we solve the same problem by using the Lie algebraic approach of quasi-exact solvability and obtain the exact solutions through the $sl(2)$ algebraization. A differential equation is said to be QES if it is an element of the universal enveloping algebra of a finite-dimensional QES Lie algebra of differential operators [10]. In one dimension, the Lie algebra $sl(2)$ is the only algebra of differential operators with finite-dimensional representations [11]. The usual realization of the $sl(2)$ Lie algebra is given by the following differential operators [10,11]

$$J_n^+=-z^2\frac{d}{dz}+nz,$$
$$J_n^0=z\frac{d}{dz}-\frac{n}{2}, \qquad (21)$$
$$J_n^-=\frac{d}{dz},$$

which leave invariant the finite-dimensional space

$$P_{n+1}=\langle 1,z,z^2,\ldots,z^n\rangle. \qquad (22)$$

Therefore, the most general one-dimensional second order QES differential equation can be expressed as a quadratic combination of the $sl(2)$ generators as

$$H = \sum_{a,b=0,\pm} C_{ab} J_n^a J_n^b + \sum_{a=0,\pm} C_a J_n^a + C, \qquad (23)$$

$$C, C_a, C_{ab} \in \mathbb{R},$$

which clearly preserves the ($n+1$)-dimensional representation space (22). In general, by means of appropriate transformations, the operator $H$ can always be reduced to a Schrödinger-type operator. Conversely, it is not difficult to verify that Eq. (5) can be expressed as a special case of the Lie algebraic form (23) as

$$H\phi(z) = 0,$$

$$H = -J_n^+ J_n^0 + (\lambda+1)J_n^+ J_n^- + aJ_n^0 J_n^- + \left((1-\frac{3n}{2}) - (\gamma+\delta+\varepsilon)\right) J_n^+ \qquad (24)$$

$$-\left((\lambda\gamma+\gamma+\lambda\delta+\varepsilon) + n(\lambda+1)\right) J_n^0 + \left(\lambda\gamma + \frac{n\lambda}{2}\right) J_n^- + \left(-\sigma - \frac{n(\lambda\gamma+\gamma+\lambda\delta+\varepsilon)}{2} - \frac{n^2(\lambda+1)}{2}\right),$$

if the following constraint (the condition of quasi-exact solvability) is satisfied

$$\frac{1}{16}\left(3 + \sqrt{1-4v_1} - 2\sqrt{1+\frac{v_3}{1+g}}\right)^2 + \frac{E}{4} = n - n(\gamma+\delta+\varepsilon) - n^2. \qquad (25)$$

As can be seen, this relation is the same as the energy relation of the BAM given in Eq. (10). Hence, as a result of Eq. (24), the operator $H$ preserves the ($n+1$)-dimensional invariant subspace $\phi(z) = \sum_{m=0}^{n} a_m z^m$ spanned by the basis $\langle 1, z, z^2, ..., z^n \rangle$ and thereby using the representation theory of $sl(2)$, we can determine the solutions of the first $n+1$ states exactly. Accordingly, the Eq. (24) can be represented as the following matrix equation

$$\begin{pmatrix} -\sigma & \lambda\gamma & 0 & 0 & 0 & 0 \\ (n-n^2)-n\xi_1 & -\sigma+\xi_2 & \ddots & \ddots & 0 & \vdots \\ 0 & \ddots & \ddots & \ddots & \ddots & 0 \\ 0 & 0 & \ddots & \ddots & \ddots & 0 \\ \vdots & 0 & 0 & \ddots & \ddots & n(n-1)\lambda + n\lambda\gamma \\ 0 & \cdots & \cdots & 0 & (2-2n)-\xi_1 & n(n-1)\xi_3 + n\xi_2 - \sigma \end{pmatrix} \begin{pmatrix} a_0 \\ a_1 \\ \vdots \\ \vdots \\ a_{n-1} \\ a_n \end{pmatrix} = 0, \qquad (26)$$

where

$$\xi_1 = (\gamma+\delta+\varepsilon),$$
$$\xi_2 = -(\lambda\gamma+\gamma+\lambda\delta+\varepsilon), \qquad (27)$$
$$\xi_3 = -(\lambda+1).$$

For a nontrivial solution to exist, the determinant of the coefficients matrix must vanish which results in certain relations between the potential parameters, which will be discussed in detail later. Also, from Eq. (4), the wave function is as

$$\psi_n(x) = (\cosh(x))^{\frac{1}{2}\left(1+\sqrt{1-4v_1}\right)} (1+g\cosh^2(x))^{\frac{1}{2}\left(1-\sqrt{1+\frac{v_3}{1+g}}\right)} \sum_{m=0}^{n} a_m \left(-\sinh^2(x)\right)^m, \qquad (28)$$

where the expansion coefficients $a_m$ satisfy the following three-term recurrence relation

$$a_m = \frac{-\left((m-1)(m-2)\xi_3 + (m-1)\xi_2 - \sigma\right)a_{m-1} - \left((6-6m)-3\xi_1\right)a_{m-2}}{\left(m(m-1)\lambda + m\lambda\gamma\right)}, \qquad (29)$$

with boundary conditions $a_{-1} = a_{-2} = 0$ and $a_{n+1} = 0$. Therefore, we have succeeded in finding the general exact expressions for the energy, wavefunction and the relation between the potential parameters. In other words, we can quickly obtain the exact solutions of the QES DWP for any arbitrary $n$ from Eqs. (25), (26) and (28). In the following, in order to explain the method in more detail and also to compare these results with those obtained by the BAM in the previous section, we study the first three states. For $n = 0$, from Eq. (25), the energy equation is given by

$$\frac{1}{16}\left(3 + \sqrt{1-4v_1} - 2\sqrt{1+\frac{v_3}{1+g}}\right)^2 + \frac{E}{4} = 0, \qquad (30)$$

and the restriction on the potential parameters is obtained from Eq. (26) as

$$\frac{\lambda}{4}\left(\frac{1}{2} + \frac{1}{2}\sqrt{1-4v_1} + \frac{1}{\lambda} - \frac{1}{\lambda}\sqrt{1+\frac{v_3}{1+g}} - v_1 - \frac{v_2}{g\lambda} - \frac{v_3}{g^2\lambda^2} + E\right) = 0. \qquad (31)$$

The energy eigenvalues of the first excited state associated with $n=1$ are given as

$$\frac{1}{16}\left(3 + \sqrt{1-4v_1} - 2\sqrt{1+\frac{v_3}{1+g}}\right)^2 + \frac{E}{4} = -(\gamma + \delta + \varepsilon), \qquad (32)$$

where the potential parameters satisfy the constraint

$$\begin{aligned}
&\left(-300E + 300v_1 + 240v_2 + 192v_3 - 250\sqrt{1-4v_1} - 970\right)\sqrt{25+20v_3} \\
&+ \left(3750E - 3750v_1 - 3000v_2 - 2400v_3 + 8125\right)\sqrt{1-4v_1} \\
&+ \left(-2500E + 2000v_2 + 1600v_3 - 17000\right)v_1 + \left(-2000E + 1280v_3 - 8600\right)v_2 \\
&+ \left(-1600E - 6680\right)v_3 + 1250E^2 + 10750E + 13725 + 1250v_1^2 + 800v_2^2 + 512v_3^2 = 0.
\end{aligned} \qquad (33)$$

Likewise, for the second excited state $n=2$, we have

$$\frac{1}{16}\left(3 + \sqrt{1-4v_1} - 2\sqrt{1+\frac{v_3}{1+g}}\right)^2 + \frac{E}{4} = -2(\gamma + \delta + \varepsilon) - 2, \qquad (34)$$

with the following constraints on the potential parameters

$$\begin{aligned}
&-\frac{125}{64}E^3 \\
&+\left(\frac{375}{64}v_1 - \frac{8325}{128} + \frac{75}{64}\sqrt{25+20v_3} + \frac{75}{16}v_2 + \frac{15}{4}v_3 - \frac{1875}{128}\sqrt{1-4v_1}\right)E^2 \\
&+\frac{E}{16000}(-150000v_2 - 120000v_3 + 3925000)v_1 \\
&+\frac{E}{16000}(375000v_2 + 468750v_1 + 300000v_3 - 3903125)\sqrt{1-4v_1} \\
&+\frac{E}{16000}\left(-30000v_2 - 37500v_1 - 24000v_3 + 376250 + 73750\sqrt{1-4v_1}\right)\sqrt{25+20v_3} \\
&+\left(-\frac{15}{4}v_2^2 + \frac{1}{16000}(-96000v_3 + 1665000)v_2 - \frac{375}{64}v_1^2 + \frac{1273}{16}v_3 - \frac{12}{5}v_3^2 - \frac{72365}{128}\right)E \\
&-\frac{32895}{32} + \frac{1}{16000}(30720v_3^2 - 1018400v_3 + 7236500)v_2 \qquad (35)\\
&+\frac{\sqrt{25+20v_3}}{16000}\left((426625 - 73750v_1 - 59000v_2 - 47200v_3)\sqrt{1-4v_1} + 12000v_2^2 + 18750v_1^2\right) \\
&+\frac{1}{16000}\left((30000v_2 + 24000v_3 - 545000)v_1 + 7680v_3^2 - 239000v_3\right)\sqrt{25+20v_3} \\
&+\frac{1}{16000}((19200v_3 - 301000)v_2 + 1174625)\sqrt{25+20v_3} \\
&+\frac{1}{16000}\left(-234375v_1^2 + (-375000v_2 - 300000v_3 + 4606250)v_1 - 96000v_3^2\right)\sqrt{1-4v_1} \\
&+\frac{1}{16000}\left(-150000v_2^2 + (-240000v_3 + 3122500)v_2 + 2430500v_3 - 9027500\right)\sqrt{1-4v_1} \\
&+\frac{53711}{160}v_3 + \frac{64}{125}v_3^3 + \frac{125}{64}v_1^3 + \frac{1}{16000}(75000v_2 + 60000v_3 - 2884375)v_1^2 + v_2^3 \\
&+\frac{1}{16000}\left(60000v_2^2 + (96000v_3 - 3140000)v_2 + 38400v_3^2 - 2453000v_3 + 17311250\right)v_1 \\
&-\frac{607}{25}v_3^2 + \frac{1}{16000}(38400v_3 - 666000)v_2^2 = 0.
\end{aligned}$$

Our numerical results for the first four states are reported and compared in table 1.

## 4. Discussion and numerical results

As shown in Eqs. (11) and (26), the three potential parameters $v_1$, $v_2$ and $v_3$ are not independent of one another, rather they are connected by certain relations. For the present problem, from Eq. (10), the parameters $v_1$ and $v_3$ must preserve the following conditions

$$\begin{aligned} v_1 &< \frac{1}{4}, \\ v_3 &> -(1+g), \end{aligned} \qquad (36)$$

to obtain real energy eigenvalues. The third parameter $v_2$, can then be determined from Eq. (26). In the following, some of our numerical results for the energy eigenvalues and the allowed values of potential parameters are reported and compared in table. 1. For comparison purpose, we use the parameter set given in Ref. [17]. It is observed that our results are identical with those obtained in Ref. [17] via a different method. Also, in Fig. 1, the variations of energy as a function of $n$ are plotted for the allowed values of the potential parameters.

**Table 1.** Solutions of the first four states for the QES DWP with $v_1 = 0.09$, $v_3 = 10$ and $g = 1/4$.

| $n$ | Energy (BAM) Eq. (10) | Energy (QES) Eq. (25) | Energy Ref. [17] | $v_2$ (BAM) Eq. (11) | $v_2$ (QES) Eq. (26) | $v_2$ Ref. [17] |
|---|---|---|---|---|---|---|
| 0 | −1.21 | −1.21 | −1.21 | −9 | −9 | −9 |
| 1 | −0.81 | −0.81 | - | −8.531128900 <br> −0.4688711260 | −8.531128874 <br> −0.4688711258 | - |
| 2 | −8.41 | −8.41 | - | −9 <br> 8.471121770 <br> −17.47112177 | −9 <br> 8.471121770 <br> −17.47112177 | - |
| 3 | −24.01 | −24.01 | - | −9 <br> −26.41460699 <br> 17.41460700 | −9 <br> −36 <br> −26.41460700 <br> 17.41460700 | - |

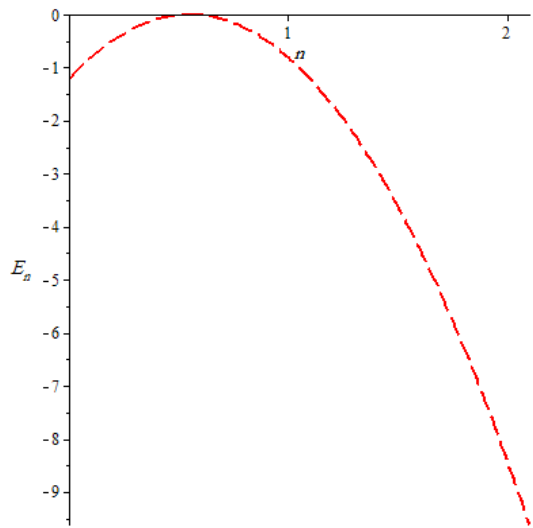

**Fig. 1.** The variations of energy versus $n$ with $v_1 = 0.09$, $v_3 = 10$ and $g = \dfrac{1}{4}$.

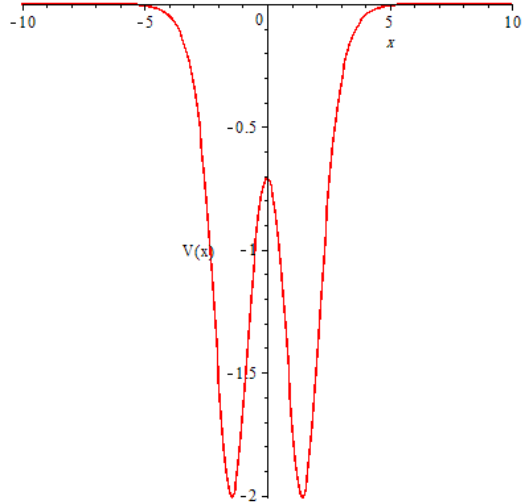

**Fig. 2.** Plot of the QES DWP as a function of $x$ for $v_1 = 0.09$, $v_2 = -9$ and $v_3 = 10$.

## 5. Conclusions

Using the Bethe ansatz method, we have solved the Schrödinger equation for a new QES DWP and obtained the general exact expressions for the energies and the corresponding wave functions as well as the allowed values of the potential parameters in terms of the roots of the Bethe ansatz equations. In addition, we have solved the same problem using the Lie algebraic approach within the framework of quasi-exact solvability and obtained the exact solutions using the representation theory of $sl(2)$ Lie algebra. Also, we have reported some numerical results and shown that the results are in good agreement with each other and with those obtained previously by using a different method.